\documentclass[12pt,preprint]{aastex}

\shorttitle{Simulations of chromospheric hard X-ray source sizes}
\shortauthors{Battaglia et al.}
\begin{document}

\title{Numerical simulations of chromospheric hard X-ray source sizes in solar flares}

\author{M. Battaglia \altaffilmark{1,2}, E. P. Kontar \altaffilmark{1}, L. Fletcher \altaffilmark{1} \and A. L. MacKinnon\altaffilmark{1}}
\affil{SUPA, School of Physics and Astronomy, University of Glasgow, G12 8QQ, UK}
\affil{Institute of 4D Technologies, School of Engineering, University of Applied Sciences and Arts Northwestern Switzerland, 5210 Windisch, Switzerland}
\email{e-mail: marina.battaglia@fhnw.ch}

\begin{abstract}
X-ray observations are a powerful diagnostic tool for transport, acceleration, and heating of electrons in solar flares. Height and size measurements
of X-ray footpoints sources can be used to determine the chromospheric density and constrain the parameters of magnetic field convergence and electron pitch-angle evolution.
We investigate the influence of the chromospheric density,
magnetic mirroring and collisional pitch-angle scattering on the size of X-ray sources.
The time-independent Fokker-Planck equation for electron transport is solved  numerically
and analytically to find the electron distribution as a function of height above the photosphere.
From this distribution, the expected X-ray flux as a function of height, its peak height and full width
at half maximum are calculated and compared with RHESSI observations. A purely instrumental explanation for the observed source size was ruled out by using simulated RHESSI images.
We find that magnetic mirroring and collisional pitch-angle scattering tend to change the electron flux such that electrons
are stopped higher in the atmosphere compared with the simple case with collisional energy loss only.
However, the resulting X-ray flux is dominated by the density structure in the chromosphere and only marginal increases in source width are found.
Very high loop densities ($>10^{11}\mathrm{cm^{-3}}$) could explain the observed sizes at higher energies,
but are unrealistic and would result in no footpoint emission below about 40 keV, contrary to observations. We conclude that within a monolithic density model the vertical sizes are given mostly by the density scale-height and are predicted smaller than the RHESSI results show.
\end{abstract}


\maketitle

\section{Introduction}

The transport of flare-accelerated electrons and the generation
of hard X-ray (HXR) emission in the solar atmosphere is one of the most important
and widely used diagnostics of flare accelerated electrons. These supra-thermal particles precipitate
along the field lines of a magnetic loop from the acceleration site towards the denser regions of the
chromosphere. They undergo Coulomb collisions with electrons and ions in the ambient plasma, and can be observed via
their bremsstrahlung emission. The bulk of the observed HXR emission comes from the footpoints of magnetic structures
where the density is high and electrons lose their energy completely, with the electron stopping
location (depth) determined by the initial electron energy and the ambient density.

Using this energy dependency in combination with observations from RHESSI \citep{Li02}, it has become possible
to characterize the structure of X-ray sources.
Assuming collisional transport, the chromospheric density could be inferred
\citep{As02, Ko08, Koet10, Sa10, Ba11a}.
These observations suggest that the bulk of HXRs in the
range 30-100~keV is produced at heights of $\sim 0.7-1.2$ Mm. The density scale heights generally
agree well with hydrostatic chromospheric density models such as \citep[e.g.][]{Ver81}. However, using RHESSI visibility techniques \citep{Hur02,
Sc07}, it is also possible to infer the energy-dependent source sizes with better than arcsecond accuracy
as shown by \citet{Ba11a,Ko08, Koet10}.
The source sizes in the direction along the magnetic field inferred by
these authors are in the range from 2 arcsec to 6 arcsec \citep{Ba11a} which is up to a factor of 4
larger than what would be expected from collisional transport in the same density profile.
To explain this discrepancy, \citet{Koet10} suggested that the electrons propagate
in a multi-thread loop with different density profiles along each thread,
so that the X-ray source positions are the same as in the case of collisional transport in a single density loop,
but the vertical size is enlarged. However, even within a single monolithic
loop, \citet{Ba11a} indicated a number of processes which might
increase the vertical extent of HXR sources.

To investigate those effects quantitatively one has to solve the equations for electron transport in the footpoints.
The electron transport problem has been considered in the past by several authors to investigate various aspects 
of electron transport, electron trapping and energy losses in the solar atmosphere. Both semi-analytical and full numerical 
solutions can be found  in the literature.
\citet{Le81,1981SvA....25..215K,1982ApJ...259..341B,1990A&A...234..487M,1992A&A...253..261M}, and \citet{2000ApJ...543..438K} use numerical methods to investigate the effect of the magnetic field geometry and magnetic trapping on the electron spectrum while \citet{1991A&A...251..693M} use a test-particle method of solution, applying it to a situation without magnetic mirroring.
Many of those studies focused on the electron transport and the effect of scattering and magnetic mirroring on the electron distribution. To compare these models with actual observations, one has to go a step further and model the resulting X-ray flux, as was done by e.g. \citet{Le83}. With the observational capabilities of instruments such as Yohkoh and RHESSI it became possible to directly compare the models with observations. \citet{Pe99} investigate the conditions for formation of loop-top sources and compare predicted time-profiles of the X-ray emission with time-profiles observed by Yohkoh. \citet{2008ApJ...673..598M} use a trap-plus-precipitation model to explain RHESSI observations. While those studies all focused on spectra, \citet{Fl96} investigated the height of X-ray sources obtained from numerical simulations and compared them with Yohkoh observations, finding that they are consistent with partial electron trapping in a magnetic trap. However, there are no quantitative studies of the HXR source sizes.

In this paper, we focus on the size of HXR sources produced by non-thermal electrons ($E\gtrsim 30$~keV) in Coulomb collisions with the ambient plasma. Within the assumption
of a single monolithic loop we analyze a range of processes which could increase the vertical extent (along the magnetic field lines) 
of the sources and explain the observations. Thus we consider: i) density variations in a single loop; ii) the role of the initial pitch 
angle distribution iii) the effect of pitch-angle scattering; iv) magnetic mirroring, and v) instrumental effects related to RHESSI observations.

\section{Transport of energetic electrons in X-ray footpoints}
The evolution of an initial electron flux distribution $F_0(E_0,z)$ as a function of distance $z$ along the magnetic field lines
is described by the time independent Fokker-Planck equation \citep[e.g.][as a recent review]{2011SSRv..159..107H},
including magnetic mirroring and collisional pitch-angle scattering and energy loss:
\begin{equation} \label{fokker}
\frac{\partial F}{\partial z}-\frac{1-\mu^2}{2\mu}\frac{\mathrm{d}(\ln B)}{\mathrm{dz}}\frac{\partial F}{\partial \mu}-\frac{Kn(z)}{\mu E}\frac{\partial F}{\partial E}+\frac{n(z)K}{2E^2\mu}\frac{\partial}{\partial \mu}\left[(1-\mu^2)\frac{\partial F}{\partial \mu}\right]=-\frac{Kn(z)}{\mu E^2}F,
\end{equation}
where $n(z)$ is the plasma density at distance $z$, $K=2\pi e^4\Lambda$, with $\Lambda$ the Coulomb logarithm, $B$ is the magnetic field strength, and $\mu$ is the cosine of the electron pitch angle relative to the magnetic field direction. In the case of purely collisional energy loss in a uniform magnetic field and an injected electron power-law flux distribution $F_0(E_0)\sim E_0^{-\delta}$ this leads to:
\begin{equation}\label{sthick}
F(E,z)=F_0[E_0(E,z)]E{\sqrt{E^2+2KN(z)}}^{(-\delta -1)},
\end{equation}
where $E_0(E,z)=(E^2+2KN(z))^{1/2}$ and $N(z)=\int n(z)dz$ is the column depth. This we will refer to as ``simple thick-target'' in this paper. 
Equation \ref{fokker} can also be solved analytically by neglecting collisions but including the term for the magnetic field \citep[e.g.][]{Le81,1983ApJ...264..648Z}. However, this analytical solution is only applicable as long as the electron pitch angle 
cosine is larger than about 0.7 because of a factor $\sim 1/\mu$ that is introduced in Eq. \ref{sthick}. 
Thus, if one wants to include all physical effects one has to use test-particle simulations.
Equation \ref{fokker} can be written as a set of stochastic differential equations \citep[e.g.][]{1981SvA....25..215K,Fl96}.
Since the typical integration time necessary for reliable imaging of footpoints with RHESSI is of the order of 30 - 60 seconds, which is very large compared to the electron loop transit time and collisional loss time in the chromosphere, a time-independent treatment is a good approximation.
Therefore we consider time-independent equations along the particle path:
\begin{eqnarray}\label{stocheq}
\frac{dz}{ds}&=&\mu \\
\frac{dE}{ds }&=&-\frac{K}{E}n(z)\\
\frac{d\mu}{ds}&=&-(1-\mu^2)\frac{d\ln B}{2dz}-\frac{\mu Kn(z)}{E^2}+\left[(1-\mu^2)\frac{Kn(z)}{E^2}\right]^{1/2}W, \label{stocheq1}
\end{eqnarray}
where $z$ is the distance from the point of injection along the magnetic field lines, $s$ is the path of the electron, 
and $W(s)$ denotes a standard Wiener process. Equations \ref{stocheq}-\ref{stocheq1} are solved using a numerical scheme:
\begin{eqnarray}
z_{j+1}&=&z_j+\mu _ j\Delta s\\ \label{stocheq_num}
E_{j+1}&=&\left[E_j^2-2{K}n(z_j)\Delta s\right]^{1/2}\\
\mu _{j+1}&=& \mu_{j}-\left.\frac{1}{2}(1-\mu_{j}^2)\frac{d\ln B}{dz}\right| _{z=z_j}\Delta s
-\frac{\mu _j Kn(z_j)}{E_j^2}\Delta s+\left[(1-\mu_j^2)\frac{Kn(z_j)}{E_j^2}\Delta s\right]^{1/2}\xi , \label{stocheq_num1}
\end{eqnarray}
where $\xi$ is a random variable taken from the normal distribution $p(\xi)=1/\sqrt{2\pi}\exp{(-\xi^2/2)}$ 
for each step $\Delta s$. The scheme proved to be reliable in modelling the effect of collisional scattering on pitch angle \citep[see][]{1982ApJ...259..341B,1991A&A...251..693M}. Equations (\ref{stocheq_num}-\ref{stocheq_num1})
are solved using a power-law electron distribution
\begin{equation}\label{power-law}
F_0(E_0,h=h_{loop})\sim E_0^{-\delta}
\end{equation}
injected at the top of the loop with high energy cutoff energy at 500 keV and injection height $h_{loop}=12.5$~Mm. Equation \ref{sthick} was used
to test the simulations. Both numerical and analytical solutions result in the number of electrons as a function
of distance $z$ from the injection point, or, equivalently, as a function of height $h$ above the photosphere $F(E,h)$
where $h=h_{loop}-z$. Finally, calculating the corresponding X-ray bremsstrahlung intensity $I(\epsilon,h)$,
and determining the X-ray source vertical profiles, this is compared to observations made with RHESSI.

\subsection{Definition of size and position}
In the observations of limb events presented in \citet{Ko08,Koet10} and \citet{Ba11a},
we found the position and size by forward fitting X-ray visibilities (2D spatial Fourier components)
with a circular or elliptical Gaussian source model. The size of a source is thereby defined as the full-width
at half maximum (FWHM) of the fitted Gaussian.
\begin{equation}
I(x,y; \epsilon)=\frac{I_0(\epsilon)}{2\pi \sigma_x\sigma_y}\exp\left(-\frac{(x-x_0(\epsilon))^2}{2\sigma_x^2}-\frac{(y-y_0(\epsilon))^2}{2\sigma_y^2}\right),
\label{eq:Imodel}
\end{equation}
where $2\sqrt{2\ln2}\sigma_x$ and $2\sqrt{2\ln2}\sigma_y$ are FWHMs
of an elliptical Gaussian source in the $x$ and $y$ direction respectively.
For HXR sources observed at the solar limb, the size along the radial direction
represents the vertical FWHM size of the source, while the perpendicular size (along the solar limb)
is equivalent to the size of a HXR footpoint parallel to the solar surface.
The simulations performed here provide the electron flux as a function
of height $F(E,h)$ (one dimension) and hence the vertical extent of HXR sources.
The X-ray flux  $I(\epsilon,h)$ per unit distance is then given as:
\begin{equation} \label{photons}
I(\epsilon,h)=\frac{n(h)A(h)}{4\pi R^2}\int_\epsilon^\infty F(E,h)\sigma(\epsilon,E)\mathrm{dE}
\end{equation}
where $A(h)$ is the area of the magnetic flux tube at height $h$, $R$ is the distance Sun-Earth
and $\sigma(\epsilon,E)$ the bremsstrahlung cross-section.

\begin{figure*}[!]
\begin{center}
\resizebox{0.4\hsize}{!}{\includegraphics{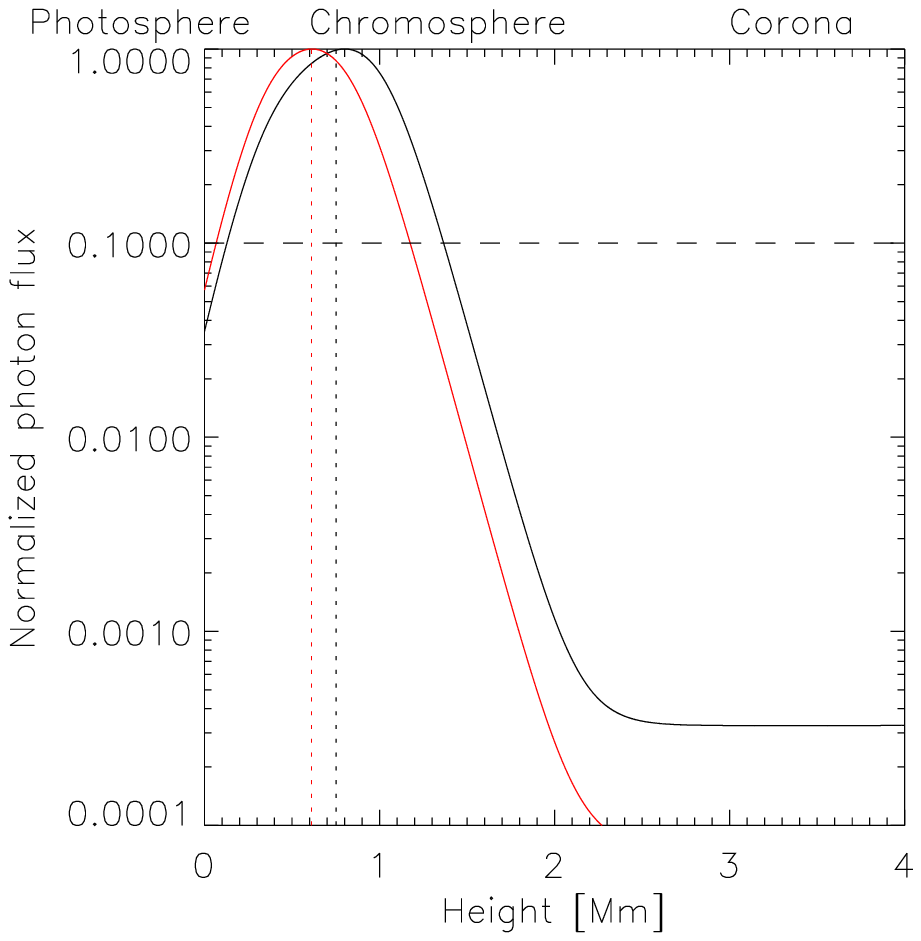}}
\resizebox{0.5\hsize}{!}{\includegraphics{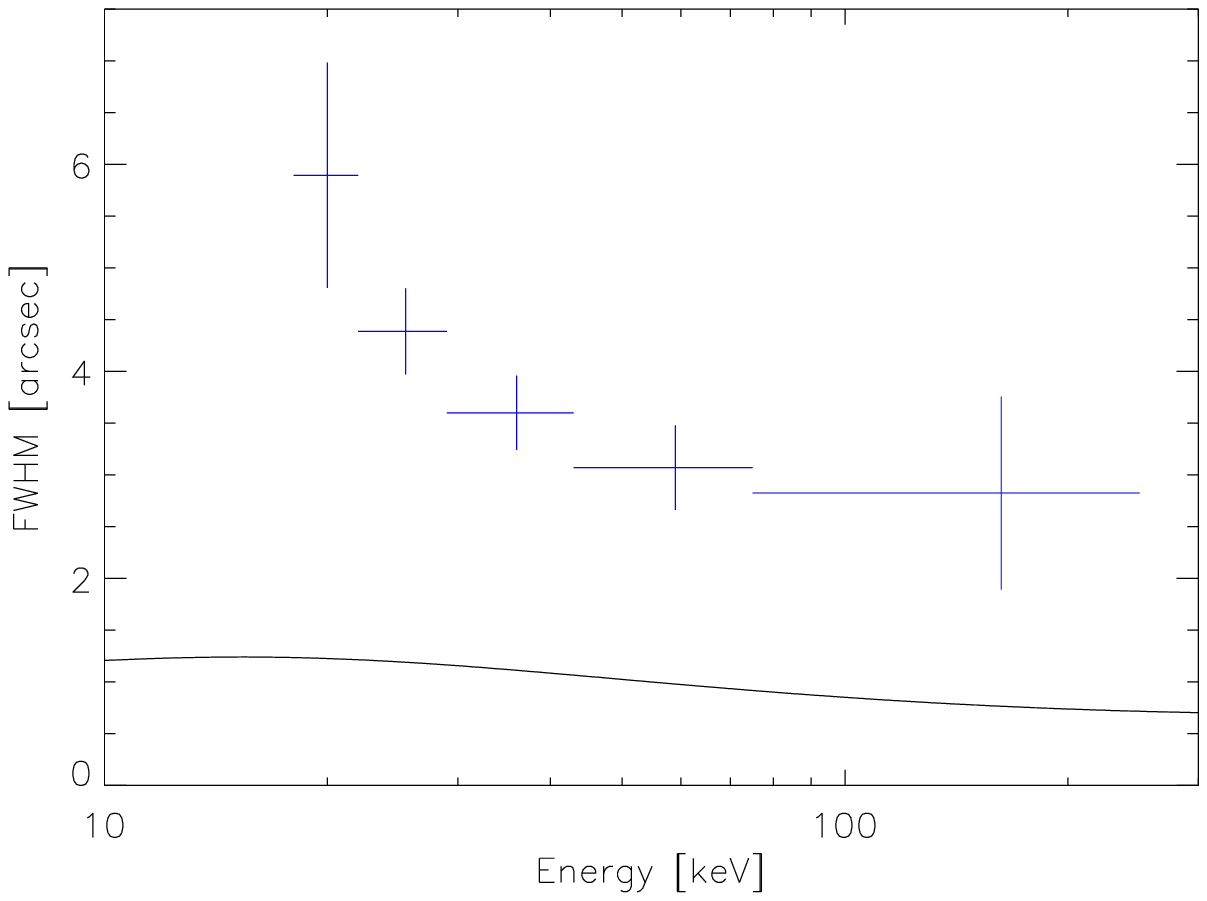}}
\end{center}
\caption {Left: Normalized photon flux as a function of height above the photosphere at 30 keV and 70 keV (red), using a density profile of $n(h)=10^{10}+1.16\times 10^{17}\times exp(-h/h_0)$ (comp. Eq. \ref{densused}) with a scale-height $h_0$=130 km. The vertical dashed lines give the first moment (maximum position) of the emission above 10 \% (horizontal dashed line). Right: Observed FWHM in the event of January 6th 2004 \citep[after][]{Koet10}. The black line is the expected source FWHM in a simple thick target, using Eqs. \ref{sthick} and \ref{photons}.}
\label{fig1}
\end{figure*}

We define the height of a source as the first moment of the X-ray flux profile and the FWHM as the second moment.
Only emission larger than 10\% of the maximum flux was used to compute the moments, to emulate
the fact that RHESSI images have a limited dynamic range \citep{Hur02}.
Figure \ref{fig1} illustrates the photon flux as a function of height for two photon energies
in the analytical solution of the simple thick-target case.
The right-hand side of the figure shows the observed FWHM in the event analyzed in \citet{Koet10}
compared with the expected FWHM in a simple thick-target model given by the analytical
solution of Eqs. (\ref{sthick},\ref{photons}) .


\section{Analytical and numerical results} \label{results}
We first investigate the effect of the chromospheric density function on the resulting source size
in the simple thick target (analytically), using Eq. \ref{sthick}. Then, the influence of the initial pitch-angle distribution, collisional
pitch-angle scattering and magnetic mirroring will be explored using test-particle simulations (Eq. \ref{stocheq} - \ref{stocheq1}).
Finally, in Section \ref{whatrhessiseas} we will discuss in more detail how the sources would
be observed by RHESSI and what results we would expect using visibility forward fitting.

\subsection{Chromospheric density structure}
\begin{figure*}[!]
\begin{center}
\resizebox{0.95\hsize}{!}{\includegraphics{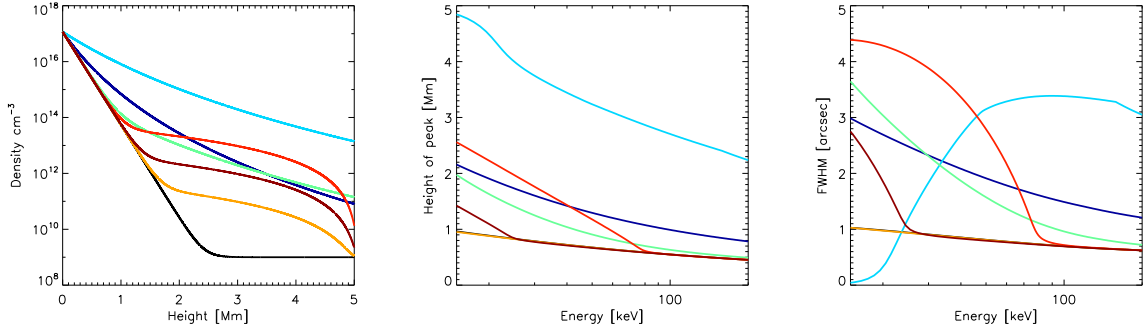}}
\end{center}
\caption {Effect of different density models on the height and FWHM of HXR sources. Left: Density models.
The black curve is an exponential density with a single scale-height of 130 km. Dark blue and light blue are $\kappa$-functions
with different scale heights. Green is an exponential density combined with a $\kappa$-function. Yellow, red and dark
red represent an exponential density combined with a $\sim h^2$ function (see text for details).
Middle: Height of peak emission as a function of energy. Right: FWHM as a function of energy. }
\label{densities}
\end{figure*}

The deposition of large amounts of energy into the chromosphere by energetic electrons could lead to a number of processes, including heating and expansion of the chromosphere. This process of chromospheric evaporation generally leads to redistribution
of plasma density in the flaring atmosphere, increasing the density of plasma in the flaring loop
\citep{1974SoPh...34..323H,An78}. At the same time, the density structure in a flaring loop
can strongly affect the source size even in the case of purely collisional energy loss. In previous work, a single scale-height
exponential density \citep{Ko08,Koet10,Ba11a}, multiple scale-height density \citep{Sa10} or a power-law density
function \citep{As02} have been investigated. Although the positions of HXR sources are in agreement
with the single scale-height exponential density profiles with scale heights of 130-200 km,
 the predicted vertical source sizes are up to a factor of 4 smaller than observed \citep{Ba11a}
(\citet{Koet10} suggested that a multi-threaded density structure with vertical strands of different density
could increase the source size significantly).
The exponential density function
\begin{equation}\label{densused}
n(h)^{exp}=n_{l}+n_{phot}\times exp(-h/h_0)
\end{equation}
is used by \citet{Ba11a} where $n_{l}$ is the (constant) loop density,  $n_{phot}=1.16\times 10^{17}$ $\rm{cm^{-3}}$ the photospheric density following \citet{Ver81}, and $h_0$ is the density scale height.
Figure \ref{densities} illustrates such an exponential density model with scale-height $h_0$=130 km, and also a density model
of the shape of a $\kappa$-function
\begin{equation}
n(h)^{\kappa}=n_{phot}\times \left(1+\frac{h}{\kappa h_0}\right)^{(-\kappa+1)},
\end{equation}
 for a $\kappa$ factor of 10 and two different scale heights $h_0=130$ km and $h_0=290$ km (dark blue and light blue curves in Fig. \ref{densities}).
 In addition, densities that combine the exponential function near the photosphere with either a quadratic function or a $\kappa$-function at larger heights were used, where the exponential plus quadratic function was given as $n(h)=n(h)^{exp}+1.16\times10^{a}(h_{max}-h)^2$, where $h_{max}$ is the loop height and $a=[10,11,12]$ (yellow, dark red, and red curves in Fig. \ref{densities}). The exponential plus $\kappa$-function was given as $n(h)=n(h)^{exp}+n(h)^{\kappa}$ (green curve in Fig. \ref{densities}).
 These latter models account for the fact that the high density in the lower chromosphere, below about 1 Mm,
 will not be affected significantly by processes such as evaporation \citep{Ma80}, while the loop density
 may be considerably different from quiet Sun density models. As Fig. \ref{densities} demonstrates,
a notable effect on the FWHM is only found if the loop density reaches extreme values of more than $10^{11}$ $\rm{cm^{-3}}$.
Such coronal densities are rather non-typically high although are observed in some flares \citep[e.g.][]{Ve06}. Such high densities
will lead to energetic electrons of energy $\sim 20$ keV and even higher being collisionally stopped in the coronal part of the loop
before reaching the chromospheric footpoints.

All these density functions result in an effective increase of the loop density compared with the single scale-height exponential
and the resulting HXR source FWHM is larger by up to a factor of 4 than in the case of an exponential density (Fig. \ref{densities}).
However, the location of the peak of the emission is also found to be a factor of 4 higher, and the bulk of the emission
below 40 keV comes from the top of the coronal loop, therefore no footpoints would be observed below 40 keV,
which is not the case in the observations \citep{Ba11a}, where footpoints are observed
at energies as low as 20 keV.

\begin{figure*}[!]
\begin{center}
\resizebox{0.95\hsize}{!}{\includegraphics{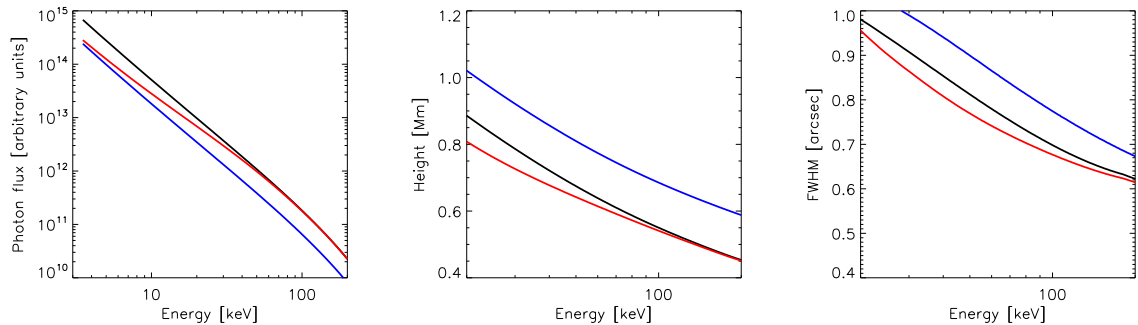}}
\end{center}
\caption {Left: Total photon spectra for completely neutral target (black), completely ionized target (blue) and ionization step-change (red). Middle and right: Height and FWHM of HXR sources as a function of energy in the three cases.  }
\label{nonuni}
\end{figure*}

 Due to the relatively modest increase of the chromospheric temperature at heights below 1~Mm, we
 assumed a neutral atmosphere. However, the plasma in the transition region and lower corona will be partially
 or completely ionized with a change in the ionization fraction at some height in the chromosphere.
The ionization state of the medium affects the Coulomb logarithm, so that $\ln \Lambda _{eH}\simeq7.1$ in neutral media
and $\ln \Lambda _{ee} \simeq 20$ in fully ionized plasma, where $\Lambda_{ee}$ and $\Lambda_{eH}$ are the electron-electron and electron-hydrogen Coulomb
logarithms. The effect of this on HXR spectra has been discussed by \citet{Br73}
and in the context of RHESSI spectroscopy by \citet{2002SoPh..210..419K}. To investigate the influence of ionization change
on the height and the FWHM of HXR sources, we  introduced an effective Coulomb logarithm
\begin{equation}
\Lambda=(\Lambda_{ee}-\Lambda_{eH})\left(x(h)+\frac{\Lambda_{eH}}{\Lambda_{ee}-\Lambda_{eH}}\right),
\end{equation}
The ionization fraction $x(h)$ was assumed to be a step function with $x(h)=1$ for $h> 0.9$ Mm
and $x(h)=0$ for $h<0.9$ Mm. The effect of non-uniform density is illustrated in Fig. \ref{nonuni}
along with a comparison with the case of constant Coulomb logarithms. The case of a more realistic atmosphere with partial ionization would lead to results within the extremes illustrated in Fig.~\ref{nonuni} and will not change the source size noticeably. Thus, the use of an idealized ionization structure is justified.

The simulations show that although different chromospheric and loop density models can increase the source size by up to a factor of 4 to fit the observations,
this will also change the height of the sources, contrary to the observations. It has to be noted that in order to produce a noticeable
change of the HXR source size, the density structure of the whole atmosphere, including the transition region and corona needs to be changed by a few orders of magnitude. This seems in contradiction with both theoretical models of the flaring atmosphere \citep[e.g.][]{2005ApJ...630..573A} and observations. 

\subsection{Pitch-angle distribution}
The height and the width of a source is also likely to depend on the initial pitch-angle distribution of the injected electron
beam. In the simple thick-target model, injection and propagation of energetic electrons is assumed parallel to the magnetic field
(i.e. cosine of initial pitch angle $\mu_0=1$), but in the other extreme case of injection perpendicular to the field ($\mu_0=0$) or strong trapping,
the electrons would lose energy near the acceleration region, the height would be constant as a function of energy,
and the FWHM would depend on the extent of the acceleration region. In an intermediate situation $0<\mu_0<1$, energetic electrons
are expected to lose energy at different heights depending on their initial pitch-angle distribution.
The effect of an initial pitch-angle distribution is investigated including collisional energy loss, but no change in the initial pitch-angle distribution is assumed,
i.e. Eq. \ref{stocheq1} is $d\mu /ds=0$. In this and the subsequent sections, an exponential density profile $n(h)=n_{l}+n_{phot}\exp(-h/h_0)$
with a scale height of 144 km and a constant Coulomb logarithm of $\Lambda=7.1$ is used. This density profile
is consistent with the source heights measured with RHESSI and the theoretical modeling of the low atmosphere
at heights $h\lesssim 1$~Mm.

\begin{figure*}[!]
\begin{center}
\resizebox{0.85\hsize}{!}{\includegraphics{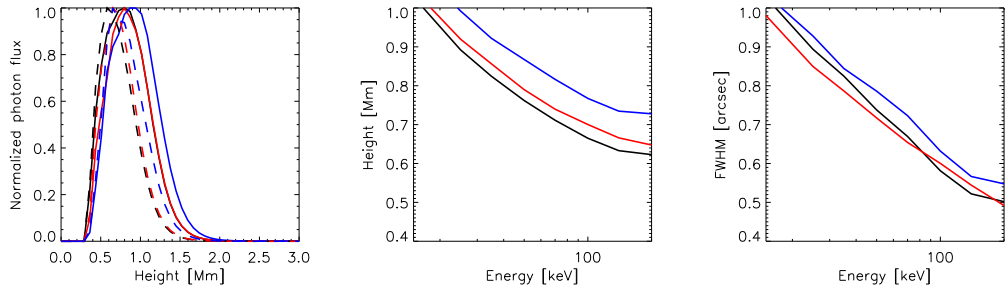}}
\end{center}
\caption {Normalized photon flux as a function of height at 30 keV and 70 keV (solid and dashed lines, left), height of maximum emission and FWHM as a function of photon energy (middle and left), for three cases of initial pitch-angle distribution: $\mu_0$=1 (black), $0.9<\mu_0\le 1$ (red), $0.1<\mu_0\le 1$ (blue). }
\label{pitchangle}
\end{figure*}

Figure \ref{pitchangle} illustrates how the initial pitch-angle distribution affects the height and the FWHM.
Three different cases are presented: $\mu_0=1$ (as a reference), $0.9<\mu_0\le 1$ uniform (strongly beamed) and $0.1<\mu_0\le 1$
uniform (broad distribution). The broad distribution leads, as expected, to a larger source size as a function of energy, as well as a larger FWHM.
However, the maximum change in both height and FWHM is about 10\% in the case of the broad distribution.
Therefore, this effect alone cannot account lead to the observed vertical sizes of HXR sources.

\subsection{Collisional pitch-angle scattering}\label{pascat}

\begin{figure*}[h!]
\begin{center}
\resizebox{0.95\hsize}{!}{\includegraphics{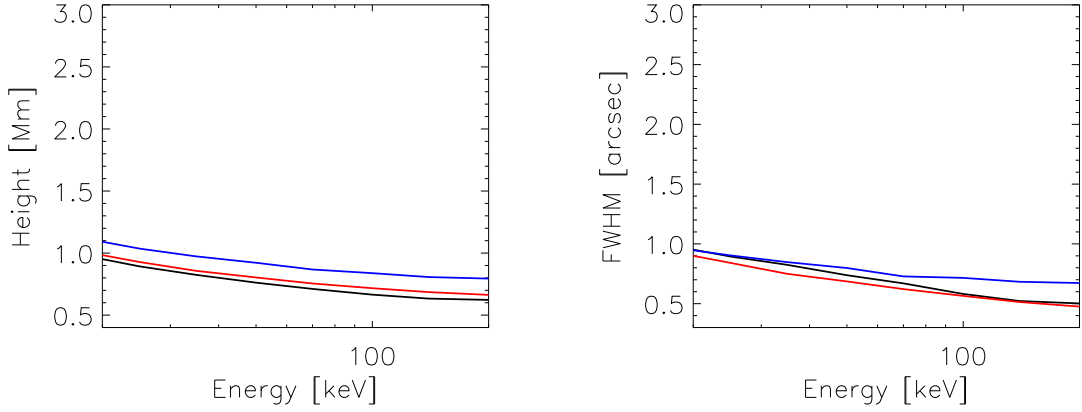}}
\end{center}
\caption {Height of maximum emission and FWHM as a function of photon energy (middle and left), for two cases of initial pitch-angle distribution: $0.9<\mu_0\le 1$ (red), $0.1<\mu_0\le 1$ (blue), and including collisional pitch-angle scattering. The black line indicates the result for collisional energy loss only.}
\label{pitchangscattering}
\end{figure*}

In collisional interactions with the ambient plasma, electrons do not only lose energy, but are pitch-angle scattered
with a similar rate. Therefore, their pitch-angle distribution changes
as the particles propagate downwards towards dense regions of the atmosphere. Figure \ref{pitchangscattering} compares
the standard case (collisional energy loss only), with the outcome of a situation with collisional pitch-angle scattering
included, i.e. Eq. \ref{stocheq1} becomes
\begin{equation}
\frac {d\mu}{ds}=-\frac{\mu Kn(z)}{E^2}+\left[(1-\mu^2)\frac{Kn(z)}{E^2}\right]^{1/2}W.
\end{equation}
In this case, the maximum emission is located up to 20 \% higher than in the no-scattering case.
However, the effect on the FWHM is small, at the level of about 5\%. As expected,
the effect is qualitatively similar to the injection of a broad initial pitch-angle distribution.

\subsection{Magnetic mirroring}

\begin{figure*}[h!]
\begin{center}
\resizebox{0.95\hsize}{!}{\includegraphics{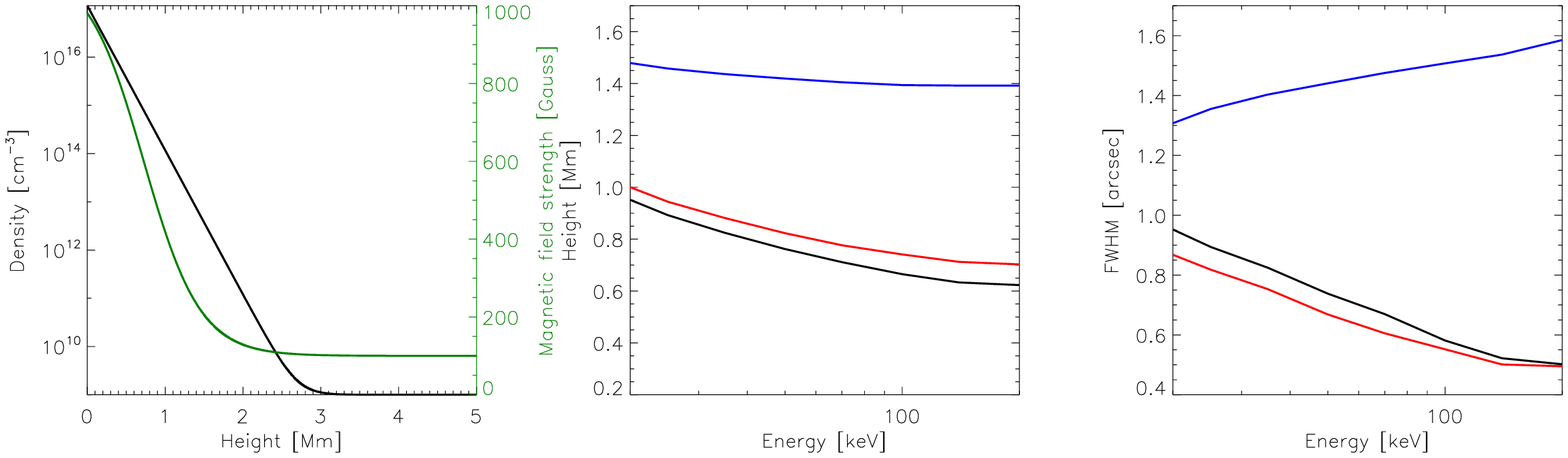}}
\end{center}
\caption {Left: Density and magnetic field strength as a function of height. Height of maximum emission and FWHM as a function of photon energy (middle and right), for two cases of initial pitch-angle distribution: $0.9<\mu_0\le 1$ (red), $0.1<\mu_0\le 1$ (blue), and including magnetic mirroring. The black line indicates the result for collisional energy loss only. $B(h)=B_0+B_1\tanh (-(h-h_M)/h_M)$ with $h_M=1$~Mm, $B_0=600$ Gauss, $B_1=500$ Gauss.  }
\label{mirroring}
\end{figure*}

A converging magnetic field at the loop footpoints also changes the pitch-angle distribution of energetic electrons, causing electrons with large pitch angle (small $\cos \theta$) to mirror upwards from the footpoints before they are collisionally
stopped. This might further contribute to an increase in source size.
In this Section we consider collisional energy loss, and pitch-angle change due to magnetic field convergence,
but not collisional pitch-angle change, i.e. Eq. \ref{stocheq1} becomes $d\mu/ds=-(1-\mu^2)\frac{d\ln B}{2dz}$. The magnetic field strength is modeled as  $B(h)=B_0+B_1 \tanh (-(h-h_M)/h_M)$, which adequately represents a converging
magnetic field in the chromosphere \citep{2011ApJ...740L..46F}. This model gives a magnetic field $B(h>>h_M)\rightarrow B_0-B_1$
at coronal heights and $B(h=0)\rightarrow B_0+B_1\tanh (1)$ at the photospheric level.
The increase in field strength relative to the ambient density is illustrated in Fig. \ref{mirroring}.
The field convergence (and the electron pitch angle) defines the depth of the mirroring point.
If the magnetic field converged higher up in the loop, the sources would be observed higher up.
In the extreme case, one could simulate a coronal source caused by electron trapping.

In the case presented in Fig. \ref{mirroring}, the source height is increased by a factor of $\sim$ 1.6,
while the FWHM increases by a factor of 1.7 - 1.3, depending on energy. Although the size increase is
larger than in the case of collisional scattering, it is still not strong enough to explain the FWHM observations.
In addition, a nearly isotropic initial distribution of electrons (Fig. \ref{mirroring}) leads to a larger source FWHM at higher energies, contrary to X-ray observations.

\subsection{Magnetic mirroring and collisional pitch-angle scattering}\label{mirrsec}

\begin{figure*}[h!]
\begin{center}
\resizebox{0.85\hsize}{!}{\includegraphics{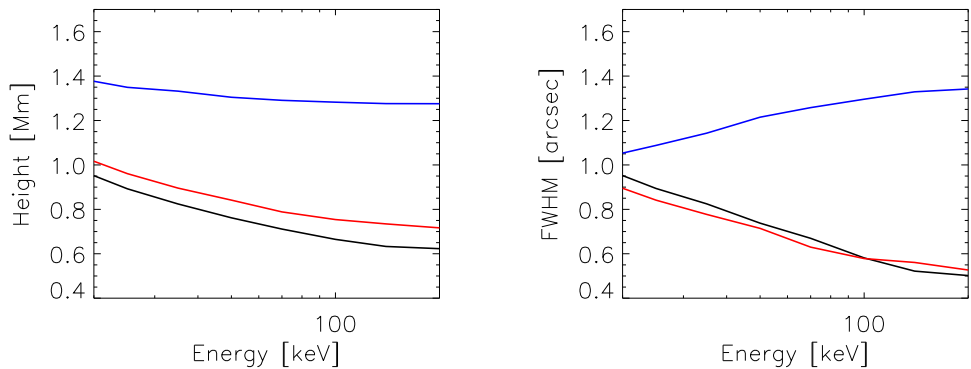}}
\end{center}
\caption {Height of maximum emission and FWHM as a function of photon energy (middle and left),
for three cases of initial pitch-angle distribution: $\mu_0$=1 (black), $0.9<\mu_0\le 1$ (red), $0.1<\mu_0\le 1$ (blue),
and including both collisional pitch-angle scattering and magnetic mirroring. }
\label{both}
\end{figure*}

Finally, the effects described in the above two sections are combined and the full Eqs. \ref{stocheq}-\ref{stocheq1}
are solved numerically (Fig. \ref{both}). Since the effect of pitch-angle scattering itself is small compared
to the effects of magnetic mirroring, this case is dominated by the effect of the magnetic field
and the result very similar to the case of magnetic mirroring (Fig. \ref{both}).

\subsection{Constructing an electron distribution as a function of height}

\begin{figure*}[h!]
\begin{center}
\resizebox{0.85\hsize}{!}{\includegraphics{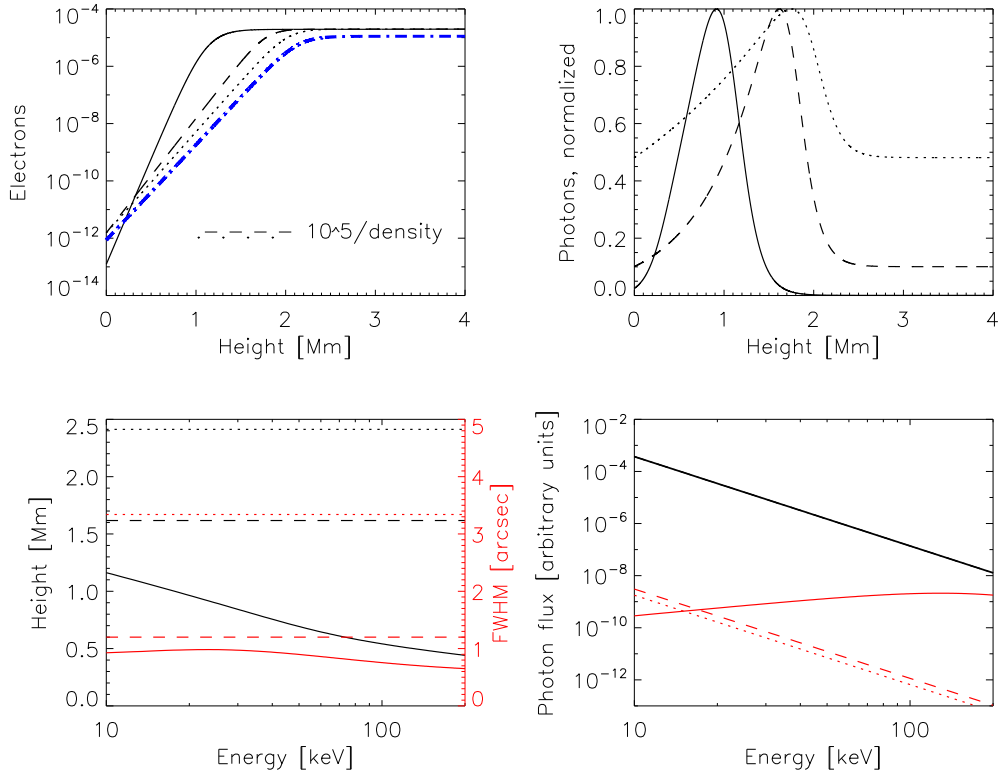}}
\end{center}
\caption {Top left: Electron flux as a function of height for 24 keV electrons. The solid line represents the simple thick target case.
The dashed and dotted lines are ``artificial'' distributions, constructed so that the slope nears that of $1/n(h)$. The blue dash-dotted line
illustrates $1/n(h)$. Top right: normalized X-ray flux. Bottom left: height and FWHM (red curves) for the three different electron
distributions where the line styles correspond to the line styles in the top left panel. Bottom right: Injected electron
spectrum (black line) and electron spectrum at a height of 0.8 Mm. }
\label{reveng}
\end{figure*}

The processes described in Sections \ref{pascat} - \ref{mirrsec} all influence the electron distribution as a function of height,
e.g. pitch-angle scattering causes electrons to be stopped higher up in the loop. However, as illustrated in Fig. \ref{pitchangscattering} - \ref{both},
this has a rather small effect on the X-ray flux profile and thus the source FWHM. The FWHM of an X-ray source
is proportional to the product of electron flux density and plasma density $I(\epsilon, h) \sim F(E,h) n(h)$. In the extreme case of $F(E,h)\sim 1/n(h)$,
the resulting $I(\epsilon, h)$ is independent of height and this constant value could extend vertically over all $h$ where $F(E,h)\sim 1/n(h)$.
We can therefore ask how $F(E,h)$ should look in order to make the product of $F(E,h)n(h)$ independent of $h$,
hence increasing the size of the X-ray source. Starting with $F(E,h)$, as found in the simple thick target case, we modified the shape
of $F(E,h)$ for every energy, so that the slope of the curve was close to $1/n(h)$, as shown in Fig. \ref{reveng} (top left),
then we computed the height of the maximum emission and FWHM. As $F(h)$ approaches $1/n(h)$, the FWHM of the resulting
X-ray flux increases up to 5 arcsec.

However, the height of the resulting HXR maximum as a function of energy is constant, as is the FWHM.  Further, the electron spectrum at low heights
is completely different from a thick target spectrum (Fig. \ref{reveng}, bottom right). Most importantly, such an electron distribution
would have to be extremely ``fine tuned'' to the ambient density.

\subsection{Instrumental effects and method}\label{whatrhessiseas}

Comparing the modeling of the main electron transport effects with observations and finding physical explanations
for the observed source sizes, we assumed that the observed difference is not entirely due to instrumental effects.
This was based on there being no modulation in the finest RHESSI grids (Grid 1 has a spatial resolution of $\sim 2.3$ arcsec and grid 2
has spatial resolution $\sim 3.9$ arcsec) in the observed events \citep{Ba11a},
indicating that the source dimensions must be of the order of $\gtrsim 4$ arcsec. Here we perform a more quantitative study of the instrumental effects,
using the simulation software developed by Richard Schwartz (private communication). The simulation software uses an arbitrary 2-dimensional
map as input and creates a corresponding calibrated event list \citep{2002SoPh..210..165S}.
The standard RHESSI imaging algorithms are then used to construct the image.
We used several source models (circular Gaussian, elliptical Gaussian) as the initial map and forward-fitted the visibilities from the corresponding
calibrated eventlist to compare the fitted FWHM with that of the original map. We find that circular Gaussians are correctly recovered
within the uncertainties, e.g. a circular Gaussian with FWHM 1 arcsec is fitted with a FWHM of $0.93 \pm 0.1$.
In the case of an elliptical Gaussian, the fitted major and minor axes tend to be larger than the original ellipse, e.g. an input elliptical source with major and minor axes of 3 and 1 arcsec respectively is fitted with 3.1 and 1.5 arcsec. Thus the fitted minor axis is 50\% larger than the input minor axis.

This last example shows that, while there are instrumental effects, especially in the case of elliptical sources,
these effects cannot entirely account for the observed source sizes.

\begin{figure*}[h!]
\begin{center}
\resizebox{0.85\hsize}{!}{\includegraphics{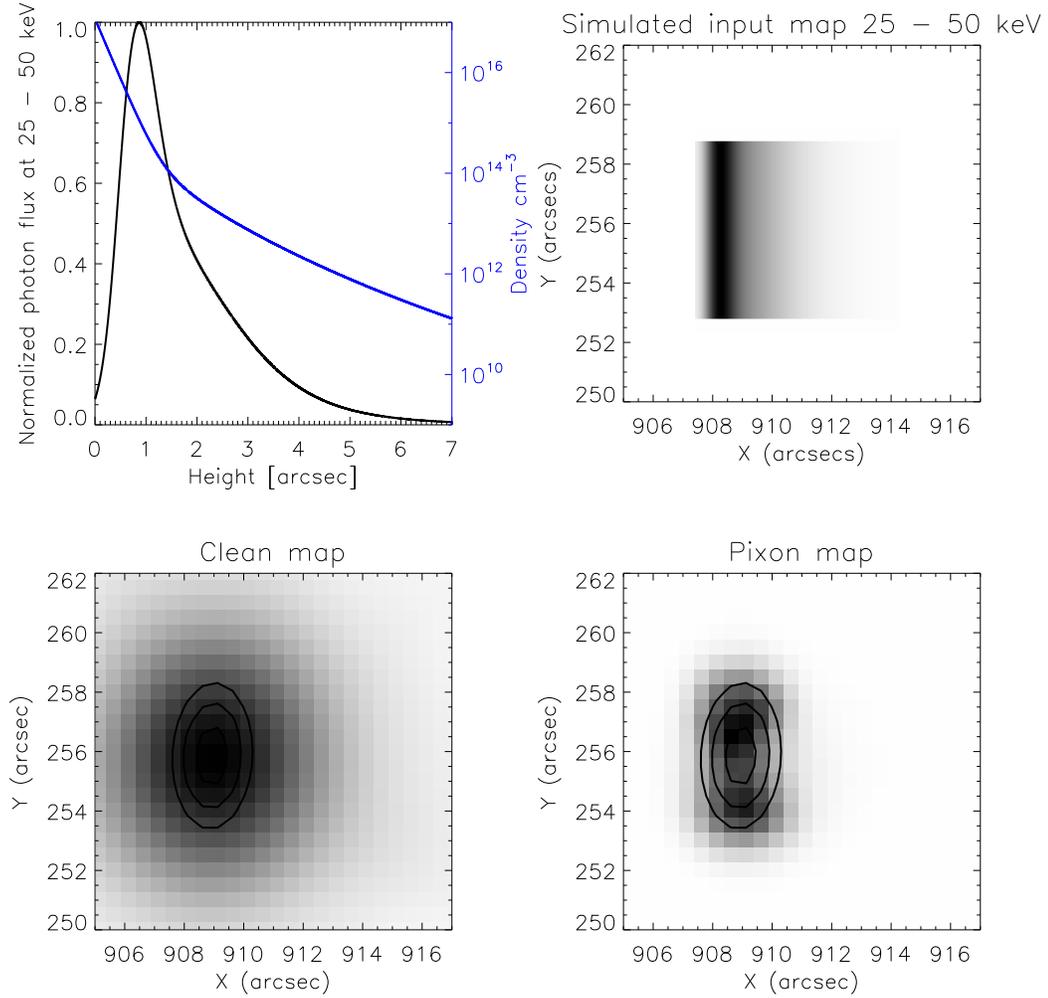}}
\end{center}
\caption {Top left: X-ray (black) profile for simulations with high loop density (blue). Top right: Simulated map of a loop with $\sim$ 6 arcsec
width and the height profile of the simulated X-ray profile. Bottom: RHESSI CLEAN and Pixon image representation
of the simulated map. The contours represent the 50\%, 70\%, and 90\% levels of the elliptical Gaussian found with visibility forward fitting. }
\label{rhessisim}
\end{figure*}

The FWHM found in the previous sections are all in terms of the second moment of the 1-dimensional height distribution of  X-rays $I(\epsilon, h)$.
However, the limited dynamic range of RHESSI influences the accuracy of the measured moments. X-ray flux in RHESSI images
which is around or less than 10\% of the brightest part of the image is dominated by an error from the brightest source. However, using the simulations
we can address the question of how such a source would be observed with RHESSI, and how the second moment found
in the simulations relates to an observed RHESSI image.  As input we used the shape of the photon-flux as a function of height
found in case of a density  $n(h)=n(h)^{exp}+n(h)^{\kappa}$ (green line in Fig. \ref{densities}) with 6 arcsec width.
This map was used as input for the simulation software and Clean and Pixon images (using software defaults and grids 1-8)
were reconstructed from the calibrated event lists. Figure \ref{rhessisim} displays the profile of the photon flux as a function of height,
the input map, and the Clean and Pixon maps. The resulting fitted FWHM are $4.9''$ and $2.7''$ for the major and minor axes.
The second moment of the 1-dimensional X-ray flux distribution are is $2.8''$ and becomes $2.1''$ if computed only for the emission
exceeding the 10\% level. Therefore, the second moment of the full distribution overestimates the size,
while the moment of the flux $> 10\%$ underestimates the size, compared to visibility forward fitting.
It has to be added that the source model was very simple and there is no unmodulated background
added to calibrated event lists.

\section{Summary and conclusions}

In the simple collisional thick-target model, both the height of the maximum HXR emission and the vertical HXR source sizes are determined by the density
scale height. RHESSI observations suggest that the HXR source positions can be well fitted with a single exponential scale-height density model, assuming a simple collisional thick target. This results in scale heights between about 130 km and 200 km
\citep{Ko08,Koet10,Ba11a}, consistent with chromospheric models \citep[e.g.][]{Ma80,Ver81}.
However, the observed HXR sizes are about 4 times larger than expected from the simple collisional transport model.
Here we have quantitatively investigated how the density profile, collisional pitch-angle scattering, magnetic mirroring, as well as instrumental effects affect the source sizes.

In \citet{Ba11a} we showed that projection effects and source motion over the RHESSI image time interval cannot account for the observed source sizes. In the present work, applying RHESSI visibility forward fitting on simulated HXR source maps we demonstrate that the source size cannot be due to instrumental effects, alone.
This leaves the physical effects of magnetic mirroring and collisional pitch-angle scattering which we investigate by solving the Fokker-Planck equation both numerically and analytically.
While pitch angle and magnetic mirroring effects change the electron flux distribution, these effects tend to increase the FWHM of the X-ray source profile by only up to a factor $1.5$, which is not enough to explain the observations.
The dominating factor that determines the X-ray source size is the atmospheric density structure.
In the case of an exponential density model with a single scale height and a constant coronal loop density of around $10^{10}$~cm$^{-3}$, the X-ray emission will originate predominantly from the region of highest density. Thus, even though the effects described above
alter the electron distribution as a function of height, emission by electrons higher up in the loop will always be faint compared to the emission from the denser chromosphere. Source sizes of around 4 arcseconds can only be achieved by unlikely loop densities of the order of $10^{13}$~cm$^{-3}$. Such high densities will also cause the HXR sources to appear at larger heights, well above typical chromospheric heights. Thus, within the traditional thick target model, the only plausible explanation for the observed HXR source sizes remains a multi-threaded density structure.

\acknowledgments
This work is supported by the Leverhulme Trust (M.B., E.P.K., L.F.) and STFC grant ST/I001808 (E.P.K., L.F., A.L.M). Financial support by the European Commission through the FP7 HESPE network (FP7-2010-SPACE-263086) is gratefully acknowledged. We thank Richard Schwartz for providing and helping with the RHESSI image simulation software and Gordon Hurford for discussions.
\bibliographystyle{apj}
\bibliography{mybib}

\end{document}